# Control of non-controllable quantum systems: A quantum control algorithm based on Grover iteration

Chen-Bin Zhang*, Dao-Yi Dong, Zong-Hai Chen

Department of Automation, University of Science and Technology of China
Hefei, 230027, P. R. China
*E-mail: Zhangcb5@mail.ustc.edu.cn

**Abstract**: A new notion of controllability, eigenstate controllability, is defined for finite-dimensional bilinear quantum mechanical systems which are neither strongly completely controllably nor completely controllable. And a quantum control algorithm based on Grover iteration is designed to perform a quantum control task of steering a system, which is eigenstate controllable but may not be (strongly) completely controllable, from an arbitrary state to a target state.

---

## 1. INTRODUCTION

In the last two decades, the issue of controllability of quantum mechanical systems has been studied by a lot of researchers from different backgrounds. Huang *et al* investigated the controllability of quantum-mechanical systems by using Nelson's analytic domain theory, Lie group and Lie algebra theory [1]. Ramakrishna *et al* studied the issue of actively controlling molecular systems in the quantum regime [2]. Schirmer *et al* obtained the sufficient conditions for complete controllability of *N*-level quantum systems subject to a single control pulse in [3,4]. Albertini and D'Alessandro analyzed the Lie algebra structure and gave out conditions of controllability for a network interacting spin 1/2 particles in a driving electro-magnetic field [6]. They also defined four different notions of controllability of physical interest for multilevel quantum mechanical systems (see [5] for details). In [7], Turinici and Rabitz presented the theoretical results on the ability to arbitrarily steer about a wavefunction for a quantum system under time-dependent external field control.

Up to now, it has been shown that the degree of controlling a quantum mechanical system depends on the dynamical Lie group of the system. The states of a quantum mechanical system could be partitioned into kinematical equivalence classes subjected to the constraint of unitary evolution of the system. If the dynamical Lie group of the system acts transitively on all these equivalence classes then the system is completely controllable or density matrix controllable and any target state in the same equivalence classes can be reached from any given initial state of the system [8].

Although many quantum systems of interest have been shown to be completely controllable, there are also others, which are either only partially controllable or not controllable at all. For these systems, the dynamical reachability of target states becomes very important in many applications. In [9], Schirmer addressed the problem by studying the action of the dynamical Lie group of pure-state-only and non-controllable quantum systems on the kinematical equivalence classes of states. In this paper, we define eigenstate controllability and present a quantum control algorithm, which is based on Grover Iteration, to consider the issue of controlling non-controllable quantum systems. With this algorithm, it is possible for us to drive an eigenstate controllable system from an arbitrary state to a desired state at will.

The paper is organized as follows: In Section 2 we describe the mathematical model we want to study and give the basic definitions. Section 3 quotes the essential part of Grover's searching algorithm that will be used in our quantum control algorithm. Section 4 presents our algorithm of controlling the eigenstate controllable quantum mechanical systems. And conclusion is presented in Section 5.

## 2. CONTROLLABILITY OF QUANTUM MECHANICAL SYSTEMS

Consider the usual quantum mechanical system described by Schrödinger equation:

$$i\hbar \frac{\partial}{\partial t}|\psi(t)\rangle = (H_0 + H_I)|\psi(t)\rangle$$

$$|\psi(0)\rangle = |\psi_0\rangle \quad (1)$$

where $|\psi\rangle$ is the state vector of complex Hilbert space. $H_0$ refers to the internal Hamiltonian and $H_I$ is the external Hamiltonian. Now, system (1) is said to be *controllable* if, given any two states $|\psi_0\rangle$ and $|\psi_d\rangle$, there exists a time interval $[0,T]$ and external Hamiltonian $H_I$ so that the system trajectory beginning at $|\psi(0)\rangle = |\psi_0\rangle$ can arrive at $|\psi(T)\rangle = |\psi_d\rangle$ under the influence of $H_I$.

In many physical situations the Schrödinger equation (1), after a truncation to a finite number of eigenstates of interest (see [2] or [7] for details), could be described as a finite-dimensional bilinear system:

$$|\dot{\psi}\rangle = (A + \sum_{i=1}^{m} u_i(t) B_i)|\psi\rangle$$

$$H_0 = i\hbar A, \quad H_I = i\hbar \sum_{i=1}^{m} u_i(t) B_i \quad (2)$$

where $|\psi\rangle$ is the state vector varying on the complex unit sphere $S_C^{n-1}$; and the matrices $A, B_1, \cdots, B_m$, are in the Lie algebra of *n*-dimensional skew-Hermitian matrices, $u(n)$. The functions $u_i(t), i = 1, 2, \cdots, m$ are time varying components of electro-magnetic fields that play the role of controls.

The solution of (2) at time *t*, is given by

$$|\psi(t)\rangle = U(t)|\psi_0\rangle \quad (3)$$

where $|\psi_0\rangle$ is the initial state and $U(t)$ is the solution of the equation

$$\dot{U}(t) = (A + \sum_{i=1}^{m} u_i(t) B_i) U(t) \quad (4)$$

with initial condition $U(0) = I_{n \times n}$. The solution $U(t)$ belongs to Lie group *U*(*n*) or *SU*(*n*) [6].

Many different notions of controllability have been defined for system (2) [3,5,7]. Here we'll give a new notion of controllability for quantum system (2). Before doing this, let us given some definitions that will be used in the rest part of this paper.

*Definition* 1. Given any $|\psi_0\rangle$ and $|\psi_d\rangle$, we say that $|\psi_d\rangle$ is *reachable* from $|\psi_0\rangle$ at time *t* if there exists an admissible control $\{u_i(t), i = 1, 2, \cdots, m\}$ such that the solution at time *t* of equation (2) is $|\psi_d\rangle$ with the initial condition $|\psi_0\rangle$. The *reachable set* from $|\psi\rangle$ at time *t*, i.e., the set of points in $S_C^{n-1}$ reachable at *t*, is denoted by $R_t(|\psi\rangle)$. In addition, the reachable set from $|\psi\rangle$ in positive time is denoted by:

$$R(|\psi\rangle) \equiv \bigcup_{t>0} R_t(|\psi\rangle)$$

Now, the controllability of system (2) could be expressed as the following definition.

*Definition* 2. System (2) is said to be *strongly completely controllable* if $R_t(|\psi\rangle) = S_C^{n-1}$ holds for all

$t>0$ and all $|\psi\rangle \in S_C^{n-1}$. If $R(|\psi\rangle) = S_C^{n-1}$ holds for all $|\psi\rangle \in S_C^{n-1}$, then the system is called *completely controllable* [1].

Note that the condition in the above definition is a little too strong and many systems of interest are non-controllable under this definition. Here we give a different definition of controllability in which the condition is a little weaker.

*Definition* 3. Suppose $|\psi_1^e\rangle, |\psi_2^e\rangle, \cdots |\psi_n^e\rangle$ are the n eigenstates of the internal Hamiltonian $H_0$, then the system (2) is called *strongly eigenstate controllable* if $\bigcup_{i=1}^{n} R_t(|\psi_i^e\rangle) = S_C^{n-1}$ for all $t>0$ and $R_t(|\psi_i^e\rangle)(i=1,2,\cdots,n)$ is called *eigenstate-from reachable set* of eigenstate $|\psi_i^e\rangle$.

If $\bigcup_{i=1}^{n} R(|\psi_i^e\rangle) = S_C^{n-1}$ holds, then the system is said to be *eigenstate controllable*.

In the definition of eigenstate controllability, we only require that for any $|\psi\rangle \in S_C^{n-1}$, there exist at least an integer $k$ ($1 \le k \le n$) such that $|\psi\rangle \in R_t(|\psi_k^e\rangle)$ (or $|\psi\rangle \in R(|\psi_k^e\rangle)$). Thus we could arrive at the following Theorem.

*Theorem* 1. If a system of form (2) is strongly completely controllable (completely controllable, respectively), it is also strongly eigenstate controllable (eigenstate controllable, respectively).

*Proof*: In fact, if the system is strongly completely controllable, then for every state $|\psi\rangle \in S_C^{n-1}$, the reachable set $R_t(|\psi\rangle)$ equals $S_C^{n-1}$. Hence $R_t(|\psi_i^e\rangle) = S_C^{n-1} (i=1,2,\cdots,n)$ holds for all $t>0$ so that $\bigcup_{i=1}^{n} R_t(|\psi_i^e\rangle) = S_C^{n-1}$ and the system is strongly eigenstate controllable. However, the converse proposition is not true.

Though the eigenstate controllability is defined for finite-dimensional systems of form (2), it can also be extended to the infinite-dimensional systems of form (1).

*Definition* 3': Suppose $|\psi_1^e\rangle, |\psi_2^e\rangle, \cdots, |\psi_n^e\rangle, \cdots$ are the eigenstates of the internal Hamiltonian $H_0$ of an infinite-dimensional system, then the system is called *strongly eigenstate controllable* if $\bigcup_{i=1}^{\infty} R_t(|\psi_i^e\rangle) = S_C^{n-1}$ for all $t>0$ and $R_t(|\psi_i^e\rangle)(i=1,2,\cdots)$ is called *eigenstate-from reachable set* of eigenstate $|\psi_i^e\rangle$.

If $\bigcup_{i=1}^{\infty} R(|\psi_i^e\rangle) = S_C^{n-1}$ holds, then the system is said to be *eigenstate controllable*.

But in this paper, we only consider the finite-dimensional eigenstate controllable systems.

Someone may wonder if the notion of eigenstate controllability is meaningful. We will see that for those systems, which are eigenstate controllable, a control law can be designed to steer the system from any initial state $|\psi_0\rangle$ to any predefined target state $|\psi_d\rangle$. The method works as follows: Suppose a system of form (2) is strongly eigenstate controllable. In order to steer an initial state $|\psi_0\rangle$ to a target state $|\psi_d\rangle$, we first find out which eigenstate-from reachable set $|\psi_d\rangle$ belongs to. Suppose $|\psi_d\rangle$ belongs to the *k*-th eigenstate-from reachable set $R_t(|\psi_k^e\rangle)$, i.e., $|\psi_d\rangle \in R_t(|\psi_k^e\rangle)$. Now if we could steer the system from $|\psi_0\rangle$ to the *k*-th eigenstate $|\psi_k^e\rangle$, then we can steer the system from $|\psi_k^e\rangle$ to $|\psi_d\rangle$ with some admissible control $\{u_i(t), i=1,2,\cdots,m\}$ by using the algorithm of [6] or [14]. Hence, the key problem has become how can we get an arbitrary state $|\psi_0\rangle$ to the *k*-th eigenstate $|\psi_k^e\rangle$.

It is known to all that if one makes a measurement on a quantum system, then the wavefunction of the system will collapse into an eigenstate with a certain probability. Suppose the wavefunction of the system is in the form of superposition of all the eigenstates:

$$|\psi\rangle = c_1|\psi_1^e\rangle + c_2|\psi_2^e\rangle + \cdots + c_n|\psi_n^e\rangle$$

$$\sum_{i=1}^{n}|c_i|^2 = 1 \qquad (5)$$

where $c_i (i=1,2,\cdots,n)$ are $n$ complex numbers. Then the probability of the wavefunction collapsing into the $k$-th eigenstate is $p_k = |c_k|^2$. Thus we can perform a measurement on a system to make the wavefunction $|\psi_0\rangle$ collapse into the $k$-th eigenstate $|\psi_k^e\rangle$ with the probability $p_k = |c_k|^2$. But we still have a problem. If the probability $p_k$ is not big enough, the wavefunction may not collapse into the $k$-th eigenstate $|\psi_k^e\rangle$ which is required if perform the measurement only once (in fact, we only have one chance). But, this problem will be solved by using Grover iteration algorithm in the next Section.

In order to apply Grover iteration algorithm, we firstly need to describe the wavefunction of $n$-dimensional complex Hilbert space in the form of $N$ qubits where $N = \text{int}(\log_2(n-1))+1$ (here the function int($x$) return the integer part of $x$). Let $\{|0\rangle, |1\rangle\}$ be an orthonormal basis for 2-dimensional complex Hilbert space. Then a two qubit system has four computational basis states denoted by $|00\rangle$, $|01\rangle$, $|10\rangle$ and $|11\rangle$. More generally, the computational basis states of $N$ qubit system can be expressed as $|x_1 x_2 \cdots x_N\rangle$ where $x_i = 0$ or $1$ ($i = 1,2,\cdots,N$). If we list these computational basis states in the order:

$$\underbrace{|00\cdots00\rangle}_{N}, \underbrace{|00\cdots01\rangle}_{N}, \cdots, \underbrace{|11\cdots10\rangle}_{N}, \underbrace{|11\cdots11\rangle}_{N}$$

and denoted them as $|1\rangle, |2\rangle, \cdots, |2^N\rangle$ for convenience, then using the first $n$ computational basis states to represent the $n$ eigenstates of a quantum system and set the coefficients of the rest $2^N$-$n$ basis states to be zero, the wavefunction in (5) could be expressed as a superposition of form:

$$|\psi\rangle = a_1|1\rangle + a_2|2\rangle + \cdots + a_n|n\rangle$$

$$+ a_{n+1}|n+1\rangle + \cdots + a_{2^N}|2^N\rangle \qquad (6)$$

where

$$a_i = \begin{cases} c_i & i=1,2,\cdots,n \\ 0 & i=n+1, n+2, \cdots, 2^N \end{cases}. \qquad (7)$$

Also for convenience, formula (6) could be expressed as

$$|\psi\rangle = \sum_{i=1}^{2^N} a_i |i\rangle \qquad (8)$$

with $\sum_{i=1}^{2^N}|a_i|^2 = 1$.

This representation may be looked as an analogy to the classical discretization of a continuous system.

## 3. GROVER'S ITERATION ALGORITHM

In this section, we only present the Grover's quantum searching algorithm in a fashion adapted to our requirements. The reader should consult the original work of Grover [10,11] for more details.

At first, we prepare a state

$$|s\rangle = \frac{1}{\sqrt{2^N}} \sum_{i=1}^{2^N} |i\rangle \qquad (9)$$

which is the equally weighted superposition of all computational basis states. This can be done by applying the Hadamard transformation to each qubit of the state $|00\cdots00\rangle$ (see [12]). Then we construct a reflection transform

$$U_s = 2|s\rangle\langle s| - I \qquad (10)$$

which preserves $|s\rangle$, but flips the sign of any vector orthogonal to $|s\rangle$. Geometrically, when $U_s$ acts on an arbitrary vector, it preserves the component along $|s\rangle$ and flips the component in the hyperplane orthogonal to $|s\rangle$. This could be understood as follows. If the system is in an arbitrary state

$$|\psi\rangle = \sum_{i=1}^{2^N} a_i |i\rangle,$$

then its inner product with $|s\rangle$ is

$$\langle s|\psi\rangle = \frac{1}{\sqrt{2^N}} \sum_{i=1}^{2^N} a_i = \sqrt{2^N} \langle a \rangle \qquad (11)$$

where

$$\langle a \rangle = \frac{1}{2^N} \sum_{i=1}^{2^N} a_i \qquad (12)$$

is the mean of the amplitude. Then if apply $U_s$ to $|\psi\rangle$, we get

$$U_s |\psi\rangle = 2|s\rangle\langle s|\psi\rangle - |\psi\rangle$$

$$= 2|s\rangle\sqrt{2^N}\langle a \rangle - |\psi\rangle$$

$$= 2\frac{1}{\sqrt{2^N}} \sum_{i=1}^{2^N} |i\rangle \sqrt{2^N}\langle a \rangle - \sum_{i=1}^{2^N} a_i |i\rangle$$

$$= \sum_{i=1}^{2^N} (2\langle a \rangle - a_i)|i\rangle. \qquad (13)$$

We can see that the $i$-th coefficient $a_i$ has become $2\langle a \rangle - a_i$ and can be looked as an operation of inversion about the mean value of the amplitude, i.e.,

$$a_i - \langle a \rangle \rightarrow \langle a \rangle - a_i.$$

If we change $|s\rangle$ with the $k$-th computational basis state $|k\rangle$ in (10), we get anther reflection transform

$$U_k = -2|k\rangle\langle k| + I \qquad (14)$$

and by applying to an arbitrary state $|\psi\rangle$, we obtain

$$U_k |\psi\rangle = -2|k\rangle\langle k|\psi\rangle + |\psi\rangle$$

$$= -2|k\rangle a_k + |\psi\rangle$$

$$= -2|k\rangle a_k + \sum_{i=1}^{2^N} a_i |i\rangle$$

$$= \sum_{i=1, i\neq k}^{2^N} a_i |i\rangle - a_k |k\rangle \qquad (15)$$

It is easy to see that $U_k$ only changes the sign of the amplitude of the $k$-th basis state $|k\rangle$ of $|\psi\rangle$. Thus we can form a unitary transformation [10]

$$U_G = U_s U_k \qquad (16)$$

which is called Grove iteration. From [10,11], we know that by repeatedly applying the transformation $U_G$ on $|\psi\rangle$, we can enhance the probability amplitude of the $k$-th basis state $|k\rangle$ while suppressing the amplitude of all the other states $|i \neq k\rangle$. If we iterate the transform enough times, then we can perform a measurement on the system to make the wavefunction $|\psi\rangle$ collapse into $|k\rangle$ with a probability of almost 1.

Let angle $\theta$ be defined so that $\sin^2\theta = 1/2^N$. Then from [13], we know that after applying the Grove iteration $U_G$ $j$ times on $|\psi\rangle$, the amplitude of the $k$-th basis state $|k\rangle$ will become

$$a_k^j = \sin((2j+1)\theta). \qquad (17)$$

If $j = (\pi - 2\theta)/4\theta$, then $(2j+1)\theta = \pi/2$ and $a_k^j = 1$. However, we must perform an integer number of iterations. Boyer has shown in [13] that the probability of failure is no more than $1/2^N$ if we perform the Grover iteration $\text{int}(\pi/4\theta)$ times. When $N$ is large, the probability of failure is very small. That is to say, we can use the Grover iteration $U_G$ to steer an arbitrary state $|\psi\rangle$ to the $k$-th basis state $|k\rangle$ with a high probability of $1 - O(\frac{1}{2^N})$.

## 4. QUANTUM CONTROL ALGORITHM FOR EIGENSTATE CONTROLLABLE SYSTEMS

Now, let's return to the control problem of quantum mechanical system (3) which is not completely controllable but eigenstate controllable. Suppose the state of the system is expressed in the form of (8), then we have a quantum control algorithm to steer the system from an arbitrary state $|\psi_0\rangle$ to any predefined target state $|\psi_d\rangle$ as follows.

**Quantum Control Algorithm:**

(i). Initialize the system in the state

$$|\psi_0\rangle = \sum_{i=1}^{2^N} a_i |i\rangle$$

(ii). Analyze the eigenstates-from reachable sets and find out the one, which $|\psi_d\rangle$ belongs to. If $|\psi_d\rangle$ belongs to more than one set, choose the one with the biggest absolute value of amplitude. Denote this eigenstate by $|k\rangle$;

(iii). Apply the Grove iterate $U_G$ on the system $\text{int}(\pi/4\theta)$ times, where

$$U_G = U_s U_k$$

and

$$U_s = 2|s\rangle\langle s| - I, U_k = -2|k\rangle\langle k| + I$$

$$|s\rangle = \frac{1}{\sqrt{2^N}} \sum_{i=1}^{2^N} |i\rangle.$$

(iv). Measure the system and the state of the system will collapse into the eigenstate $|k\rangle$ with a probability of $1 - O(\frac{1}{2^N})$.

(v). Use some admissible $\{u_i(t), i=1,2,\cdots,m\}$ to drive the system from the eigenstate $|k\rangle$ to the destination $|\psi_d\rangle$. This could be done by using the control algorithm of [6] or [14].

*Remark* 1. From Section 3, we know that the above algorithm will fail to work with a probability of no more than $1/2^N$. If $N$ is large enough, the control algorithm may succeed with a high probability. So this algorithm is essentially a probabilistic algorithm. On the other hand, though the algorithm is presented for an eigenstate controllable system, it is obvious that the control algorithm will also work on a completely controllable system.

*Remark* 2. The step (ii) of the above algorithm is to analyze the structure of the eigenstate-from reachable sets. It is just the knowledge about the reachable sets that make the control scheme possible. It is similar with the system analyzation of design which is most important in classical control theory.

*Remark* 3. From this algorithm, we can see that a measurement of a quantum system may also be looked as a kind of control. By using quantum measurement properly, one can make some impossible control task possible in some quantum mechanical systems.

## 5. CONCLUSION

The controllability of quantum systems using external control field has been studied before by various authors. However, the (strongly) completely controllability is a little too strong and many systems of interest are not (strongly) completely controllable. Thus, in this paper, we give a weaker definition of controllability which is called (strongly) eigenstates controllability. And for these quantum mechanical systems, which are eigenstates controllable but may not be completely controllable, we designed a quantum control algorithm based on Grover iteration to steer the system from any initial state $|\psi_0\rangle$ to any predefined target state $|\psi_d\rangle$. This algorithm is a probabilistic algorithm and will work with a probability of almost 1. But it still has the possibility of failing to work. The algorithm is defined for pure-state quantum mechanical system. How to adapt the algorithm to work for mixed state quantum mechanical systems will be our work in the future.